# Photonic biosensing via Su-Schrieffer-Heeger boundary modes

Yang Liu[1] and Jian-Hua Jiang[2, 3, 4, 1, †]

[1]School of Physical Science and Technology & Collaborative Innovation Center of Suzhou Nano Science and Technology, Soochow University, Suzhou 215006, China

[2]School of Biomedical Engineering, Division of Life Sciences and Medicine, University of Science and Technology of China, Hefei 230026, China

[3]Suzhou Institute for Advanced Research, University of Science and Technology of China, Suzhou 215123, China

[4]Deparment of Modern Physics, School of Physical Sciences, University of Science and Technology of China, Hefei 230026, China

[†]*Correspondence to jhjiang3@ustc.edu.cn (JHJ)*

**Abstract**

We propose a conceptual device for multiplexed biosensor in a photonic crystal chip based on the Su-Schrieffer-Heeger mechanism. Remarkably, the proposed biosensor can identify three distinct disease markers through a single-shot photon transmission measurement, due to the couplings among the three Su-Schrieffer-Heeger boundary modes in the photonic crystal. Our biosensor design is more robust against defects and disorders inevitable in real-life device applications than previous designs. Such robustness is invaluable for achieving efficient, reliable, and integrated biosensing based on nanophotonic systems. We further demonstrate that various combinations of disease markers can be recognized via the photon transmission spectrum, unveiling a promising route toward high performance advanced biosensing for future biomedical technology.

**Introduction**

On-chip photonic biosensing utilizing high-quality resonant cavities offers a convenient way for detecting the biomarkers relevant to medical diagnosis and biological research [1, 2]. Via the state-of-the-art lab-on-a-chip (LOC) technology [3-7], sensitive and rapid photonic biosensing can be achieved in an unprecedented way. Progresses in this direction include label-free optical biosensors [8-15], surface plasmon resonance biosensors [16-26], and other photonic biosensors [27-32] that can be used to detect chemicals and biomaterials for fundamental research and cutting-edge applications in security, food safety, health, and other fields. Once photonic biosensors are integrated on a chip, then many biomarkers can be detected through single-shot measurements, heralding a promising future for biosensing with high efficiency and capacity. However, such integration is currently achieved in a straightforward way where independent biosensors are detected in parallel optical paths. Higher-level photonic integration will considerably improve the detection capacity and efficiency of biomarkers.

Here, we propose a photonic crystal (PhC) architecture for multiplexing photonic biosensors based on the Su-Schrieffer-Heeger (SSH) boundary modes. By designing a 2D square lattice PhC with a nontrivial Zak phase [33], we create a configuration where two topological edge modes and a soliton-like interface mode emerge simultaneously in a clean photonic band gap [34]. The transmission through the PhC is then modulated by these boundary modes. Interestingly, we find that unconventional transmission spectra appear in such a setup which cannot be explained without considering the couplings among the interface modes. When biomarkers are bound to the structures at these boundary modes, the photonic transmission spectrum exhibits intriguing Fano resonances where the transmittance can vary with the amount of biomarkers attached to the biosensors. These properties offer the logic discrimination of various combinations of three different biomarkers and thus provide a scheme for multiplexed photonic biosensing with high efficiency and capacity.

**Results**

Our lab-on-a-chip (LOC) PhC biosensor uses the liquid with tissue cells as the sample to detect disease markers. The liquid sample interacts with the material (such as a dielectric pillar with a biometric surface layer) possessing biometric properties that attract and bind the targeted biomarkers. Such binding will be captured by the refractive index sensing through photon transmission measurements where the binding of the biomarker will alter the resonance frequency of the cavity mode at the binding site [35-37].

The SSH model describes a one-dimensional lattice with dimerized couplings [34]. Depending on whether the inter- or intra- unit cell coupling is stronger, the system is classified as either topological or trivial. In the topological case, edge states emerge at the edge boundaries or at the interface boundary between the trivial and topological chains. In the trivial case, there is no edge states. In PhCs, the photonic version of the SSH model can be realized by modulating the distance between the dielectric pillars, since the transverse-magnetic (TM) wave couplings between these pillars are controlled by distance. It has been shown that such photonic SSH model has been realized in both 1D and 2D PhCs [38-40]. Furthermore, in PhCs, the topological invariant for the SSH model, i.e., the Zak phase can be deduced from the parity of the photonic Bloch wavefunctions at the high-symmetry points of the Brillouin zone [33].

By combining two SSH PhCs in topological and trivial phases, we can obtain a photonic interface mode. After several such continuations, a long-chain PhC with numerous interface modes is architected. Covered by these interface modes, thickless biometric surfaces are pinned on the dielectric pillars. When disease markers touch the biometric surfaces, antigen-antibody binding reactions will be activated, causing accumulated protein-binding layers on the biometric surfaces (see the schematic in Fig. 1(a)). Such a biological reaction alters the refractive-index configuration of our PhC, and so on the mode couplings and hybridizations. Variable mode couplings and hybridizations are the underlying physical mechanisms that determine the measured output light.

**1.Design of the asymmetric edge modes**

In this work, we propose a two-dimensional PhC purely made of conventional dielectric materials. One basic cell has two rectangle silicon pillars (relative dielectric constant, $\varepsilon_r = 11.56$) to capture the low-frequency electric fields, and they are surrounded by the liquid sample ($\varepsilon_r = 1.8225$) containing tissue cells. In figure 1(b), we picture two "standard" unit cell structures, in which the length and width of the silicon pillar (magenta rectangle) are $0.4a$ and $0.2a$ to ensure equality in $x$ and $y$ directions. Here, "$a$" is the lattice constant, and "standard" means the maximum complete bandgap, $0.2653c/a - 0.3271c/a$. By oppositely pushing two silicon pillars in the $x$ direction, one can continuously drive unit cell 1 to unit cell 2, accompanied by a topological phase transition. As needed, a topological invariant, the Zak phase, determined by the evolution of wavefunctions in Brillouin zone, should be introduced to characterize such a one-dimensional topology transition. One of the simplest methods to obtain the Zak phase is by comparing the parities of the electric fields at the first band between different high-symmetry points. As depicted in the bottom of Fig. 1(b), even-parity electric fields at both $\Gamma$ and $X$ points for unit cell 1 cause a trivial Zak phase 0. Due to the equivalent $a/2$ translation bridging cell 1 and cell 2 in $x$ direction, a parity transition emerges at the unit cell 2's $X$ point, deriving a non-trivial Zak phase $\pi$. At the open boundary position, the unit cell with $\pi$ Zak phase can generate the topology-protected edge state, differing from the trivial case. Such edge modes guaranteed by nontrivial topological invariants play a vital role in our work. The protein-binding layer ($\varepsilon_r = 2.1025$) coating the pillars covered by these edge modes will modulate their electric fields, causing frequency shifts and varying transmission properties. These regulatory functions based on edge modes critically guide the design of our biosensor.

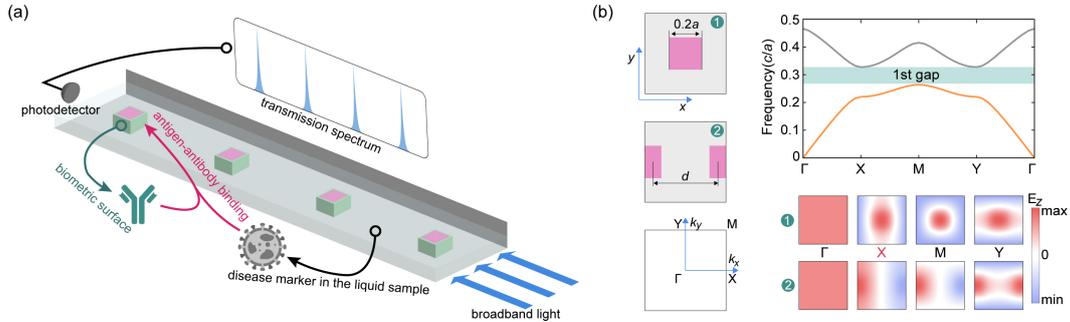

**Figure 1.** (a) Schematic of on-chip dielectric-material biosensors. Magenta squares denote the dielectric pillars. Green boxes represent the biometric surfaces (including the protein-binding layers) with antibodies to diagnose the disease antigens brought by the liquid (labeled by light blue with high transparency). A broadband light tunnels the biosensor and is then collected by a photodetector. (b) Configuration of the photonic-crystal-version SSH unit cells. The left panels are the trivial, nontrivial unit cells, and their shared Brillouin zone. $a$ means the lattice constant. $d$ indicates the distance between two silicon rectangles. $\Gamma$, X, M, and Y are the high-symmetry points for our cells. The right panels present the shared band diagram and unit cell's electric fields ($E_z$) at previous high-symmetry points. Red X labels the emergent parity reversion.

Generally, photonic crystals inspired by the well-known SSH model tend to engineer a symmetric configuration. In this work, we stack four SSH arrays whose Zak phases wave between 0 and $\pi$, like a square-wave function. Each SSH array consists of four identical cells whose length is suitable for mode couplings. The combined strip-like structure encapsulated by glasses ($\varepsilon_r = 2.25$ and the thickness is $0.5a$) exhibits an obvious asymmetry in $x$ direction (see Fig. 2(a)). There are four topological edge modes emerging at the $L_0$, L, C, and R interfaces. Additionally, a special edge mode lacking the topological protection is located at the $R_0$ interface. Edge modes bound at $L_0$ and $R_0$ interface seem to decouple with another three edge modes. We can gather the residual edge modes to form a complete-like space that we focus on. To amplify the asymmetry of structure and edge modes, the geometry parameters of these silicon pillars in four regions have been finely tuned. From left to right, the size of silicon pillars takes $0.362a * 0.181a$, $0.476a * 0.238a$, $0.4a * 0.2a$, and $0.4a * 0.2a$, respectively.

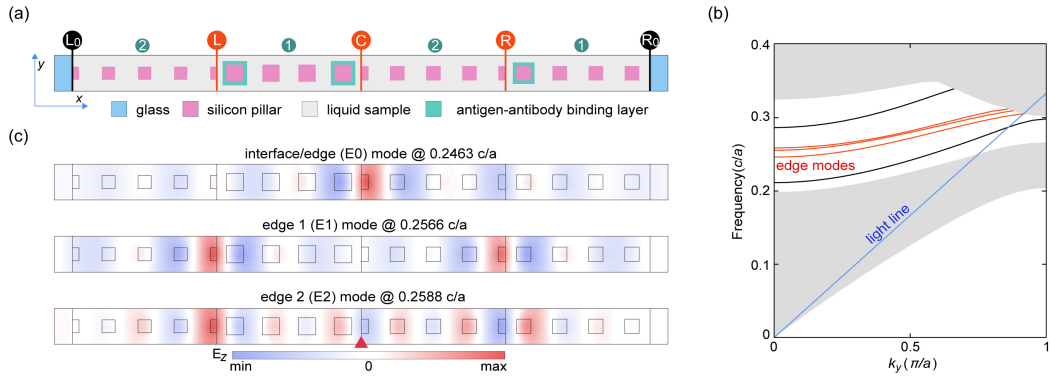

**Figure 2.** (a) Illustration of a PhC SSH strip embedded in two glass layers. In-circle L, C, and R label the interfaces between different SSH arrays. Numbers in green circle mark the topological classification of SSH arrays. Number 1 means the array is trivial, and number 2 means the nontrivial topology. In-circle $L_0$ and $R_0$ denote the interface between glass layer and SSH array. (b) Band diagram of the biosensor strip periodic in $y$ direction. Blue line is the light line of glass. In-gap red curves indicate the edge modes mainly distributed at L, C, and R interfaces. In-gap black curves represent the edge modes placed at $L_0$ and $R_0$ interfaces. (c) Field distributions $E_z$ for three in-gap edge modes that bound to L, C, and R interfaces. The tiny red triangle tips the abnormal weak electric field at the C interface for E2 mode.

Utilizing periodic arrangement, we can copy our PhC strip along $y$ direction to obtain numerous interfaces, where plentiful biometric layers can be set up to probe various diseases. Such an expanded structure can be regarded as a highly integrated biosensor. In figure 2(b), we give its band diagram. Two gray banners correspond to the first two bulk bands of unit cells. Optical modes above the blue line (the light line of glass) leak from the glass encapsulation and work as detected signals. Five in-gap curves depict edge modes stemming from five interfaces shown in Fig 2(a). Three of them labeled red are distributed at L, C, and R interfaces. Since light strikes the glass surface perpendicularly, we only show their electric field $E_z$ configurations when $k_y = 0$ (see Fig. 2(c)). Electric field of the lowest-frequency ($0.2463c/a$) edge mode (E0 mode) is mainly concentrated at the C interface, like a soliton state, with a weak antiphase field at the R interface. According to the electric field distribution, E0 mode is nearly decoupled from the other two edge modes. As a result, it can be used as an independent reference to directly detect the antigen-antibody binding reaction attached to the biometric pillar near the C interface. The remaining two edge modes, E1 and E2 modes, are a couple of symmetric- and antisymmetric-like edge modes. They are close in both frequency ($0.2566c/a$ and $0.2588c/a$) and real-space domains (mainly distributed at both L and R interfaces). One visible difference is that a weak electric field (pointed out by a red triangle) occurs near the C interface in E2 mode but not in E1 mode, which is also consistent with the result of E0 mode. Such particularity forges the high-resolution ability of E2 mode discussed in next section.

**2. Analyses of transmission spectra**

Light enters our PhC strip through the left glass layer. The forbidden gaps of four SSH arrays obstruct the light falling into them via Bragg scatter. Thus, in the bandgap shown in Fig. 2(b), only the in-gap states, five edge modes, help the light evanescently pass through the SSH arrays and then exit from the right glass layer, like the resonance transport. During this process, a maximum bandgap can offer a clear transmission background, and complex mode couplings will bring abundant transmission fingerprints. Obviously, coupling between interface modes and glass modes, controlled by the overlap of their field distributions, is the critical factor that determines the transmission properties. Each SSH array made of four cells gives a relatively even spatial distance $4a$ between interface modes, supporting the edge mode channels with both high transmittances and high-quality factors (see Fig. 4(a)). Stretching the SSH array causes lower transmittances but higher quality factors, and vice versa. In our work, four cells per array is a suitable setup for transmission spectra.

In addition, the protein-binding layer attached to the silicon pillar can affect the transmission performance through refractive index sensing. In this way, the choice of silicon pillar dressed by a biometric surface modifies the response of the transmission spectra to the antigen-antibody binding reaction. More coincident biometric surface and mode field intensity will result in more sensitive refractive sensing as well as a more pronounced transmission response. Intensity distributions for E0, E1, and E2 mode are given in Figs. 3(a-c). As can be seen from the figures, the major peaks of L, C, and R interface modes cover the extra half squares in nontrivial SSH arrays. In the following discussion, we chose the minor intensity peak of an interface mode to locate the biometric surface. The selected silicon pillars are shown as the magenta squares coated by protein-binding layers (green hollow boxes) in Fig. 2(a).

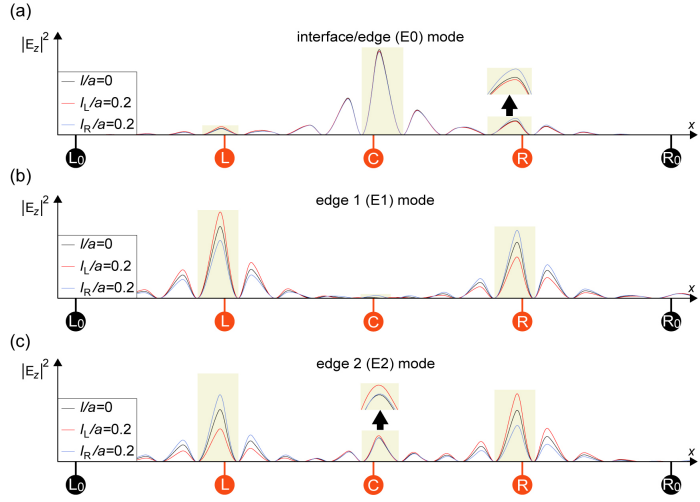

**Figure 3.** Electric field intensity diagrams $|E_z|^2$ for E0-(a), E1-(b), and E2-(c) modes. $L_0$, L, C, R, and $R_0$ label all five interfaces in our biosensor strip. Black line describes the field intensities without protein-binding reaction. Red and blue lines correspond to the L- and R-type bindings with the same binding layer thickness $l/a = 0.2$. Beige banners highlight the major intensity peak for L, C, and R interface modes.

Assume that an optical-plane wave enters our PhC strip and exits from the right glass layer. By varying its frequency, one can obtain a spectrum with respect to transmittance and frequency. Using the radio frequency module of COMSOL, we simulate the in-gap transports through E0, E1, and E2 modes. The reference transmission spectrum is shown in Fig. 4(a), where the biological reactions are suppressed. There are three transmission peaks in the frequency domain: $0.244c/a - 0.260c/a$. According to their frequency centers, one can find that, from low to high frequency, these three peaks stem from the E0, E1, and E2 edge modes, respectively. Hence, these peaks can probe the protein-binding reactions affecting edge modes.

According to the spatial locations where biological reactions occur, all protein-binding configurations can be categorized into seven types: L, R, LR, C, LC,

CR, and LCR. Here, we elaborate on L, R, LR, and C binding types in detail, and the residual types can be understood by their direct summation. The varying thicknesses of the protein-binding layers $l/a$ here imply the strength of antigen-antibody binding reaction. Firstly, figure 4(b) illustrates the transmission spectrum for L-type protein binding. Biometric silicon pillar near the L interface employs the antigen-antibody reactions to continuously accumulate the protein-binding shell. As the protein layer thickens, both the E1 and E2 peaks exhibit redshifts, with their transmittances varying oppositely. The E1 peak decrease, while the E2 peak increase. Given that the thickened protein layer ($n = 1.45$) have a higher refractive index than the liquid sample ($n = 1.35$), all transmission peaks in all binging configurations will experience redshifts as the layers thicken. The only distinction is that, their redshift levels determined by the overlap between protein layers and mode's field distribution will be different. E1 and E2 peaks have visible redshifts as they distribute near-half field intensity at the L interface. E0 peak appears to decouple from the L interface mode (or L-type binding) owing to its soliton-like field pattern. An additional consideration is the variation in transmittance. In our architecture, photons in left glass can be coupled into the SSH arrays through L and C interface modes. Photons in SSH arrays can tunnel to the right glass layer via C and R interface modes. Clearly, mode couplings govern the photon transmissions. In our PhC biosensor, asymmetric architecture induces the asymmetric edge modes, and, consequently, the asymmetric mode couplings. Figure 2(c) illustrates that L interface mode, primarily placed at the left side of its interface, is closer to the glass layer compared to the R interface mode. Thus, the coupling between left glass and L interface modes is stronger than that between R interface and right glass modes, which causes a scattering rate relation, $\gamma_{l-L} > \gamma_{R-r}$ ($l$ and $r$ mean the left and right glass). This asymmetric scattering relation, together with the asymmetric interface mode couplings and hybridizations, gives rise to the opposite variations in transmittances of E1 and E2 peaks: a decreased E1 peak and an increased E2 peak (for further details, see Appendix). When protein-binding reactions occur near the R interface, redshifts for both E1 and E2 peaks still emerge, but their transmittance trajectories are reversed, as depicted in Fig. 4(c). R-type protein binding brings an increased E1 peak and a decreased E2 peak. If protein-binding shells accumulate concurrently at L and R interfaces, the opposite transmittance variations inherited from the L- and R-type bindings counteract each other. Since the binding silicon pillar at the L interface is larger than that at R interface, the L-type protein-binding layer with the same thickness will exert relatively stronger power. As Fig. 4(d) describes, a slight decrease of E1 peak and a weak increase of E2 peak are maintained as consequences. Considering the frequency variations and the joint effects of L- and R-type bindings, the double redshifts observed in both E1 and E2 peaks are expected. Additionally, since the E0 mode distribution is not absolutely clean (such as the small intensity peaks at L and R interfaces in Fig. 3(a)), its transmission peak irresistibly accumulates a weak redshift, which is invisible in L- and R-type bindings. We now turn attention to C-type protein binding, for which the transmission spectrum is given in Fig. 4(e). In our simulation, due to the high Q and high transmittance of E0 mode, only the frequency of E0 peak can be modulated. Based on the intensity peaks at C interface in Figs. 3(a-c), it can be inferred that, as C-type binding accumulates, E0 peak redshifts dramatically, E2 peak redshifts weakly, while E1 peak appears insensitive to C-type protein binding. These are consistent with the simulation results. Finally, incorporating C-type binding with L-, R-, and LR-type bindings, we obtain the LC, CR, and LCR binding configurations, as shown in Figs. 4(f-h).

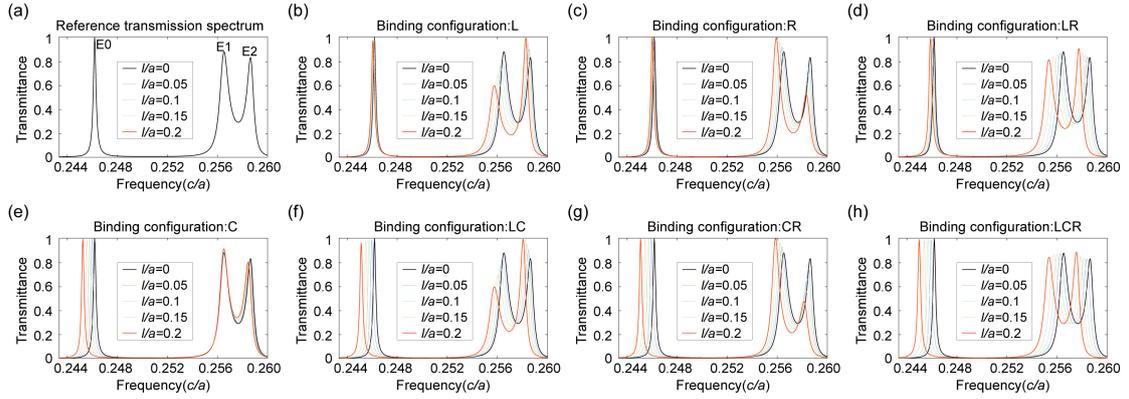

**Figure 4.** Transmission spectra for different binding configurations. (a) is the reference spectrum without protein-binding reactions. (b-h) give the transmission spectra for L-, R-, LR-, C-, LC-, CR-, and LCR-type protein binding with binding layer thickness $l/a$ varying from 0 to 0.2.

By shifting the perspective, we can observe how a transmission peak responds to seven different types of protein bindings. In addition to the invariant transmittance of E0 peak, there are five controllable observations in total, such as E0-E2 peaks' frequency shifts and E1-E2 peaks' transmittance shifts. The relevant results reorganized from figure 4 are shown in Figs. 5(a-e). As Fig. 5(a) describes, E0 peak's frequency can effectively distinguish the signals w/ or w/o C-type binding. Consistent with previous discussion, E0 mode serves as an excellent tool to analyze C-type protein binding, as stated by Step 1 in Fig. 5(f). For E1 peak, its frequency response (see Fig. 5(b)) clearly separates the signals w/o L/R/LR-, or w/ L/R-, or w/ LR-type binding. While it fails to distinguish L- and R-type bindings. Fortunately, both E1 and

E2's transmittance shifts (Figs. 5(d-e)) compensate for this shortage, as they can easily diverge L- and R-type bindings with opposite response trends. In detail, they can even trigger the difference between the signals w/ or w/o LR-type binding, as denoted by the red bidirectional arrows in Figs. 5(d-e). Obviously, E1 peak's frequency shift together with E1 or E2 peak's transmittance shift are sufficient to distinguish all binding configurations except for cases w/ or w/o C-type protein binding. These are expounded by the Step 2 in Fig. 5(f). Via Step 1 and then Step 2, one can analyze all binding cases even in a low-density antigen environment. Finally, we analyze the most unique response: E2 peak's frequency shift (Fig. 5(c)). As shown in the intensity diagram Fig. 3(c), E2 mode is near-evenly distributed at L, C, and R interfaces. Field pattern determines the sensing ability of biometric surfaces at L, C, and R interfaces. Therefore, E2 peak's frequency shift exhibits a unique divergence depending on the number of active binding positions, which gives it extraordinary discernment. Thus, in a high-density antigen environment, E2 peak's frequency alone has a high-resolution ability to distinguish all binding configurations. Resolution abilities of five variable observations are summarized in Fig. 5(g).

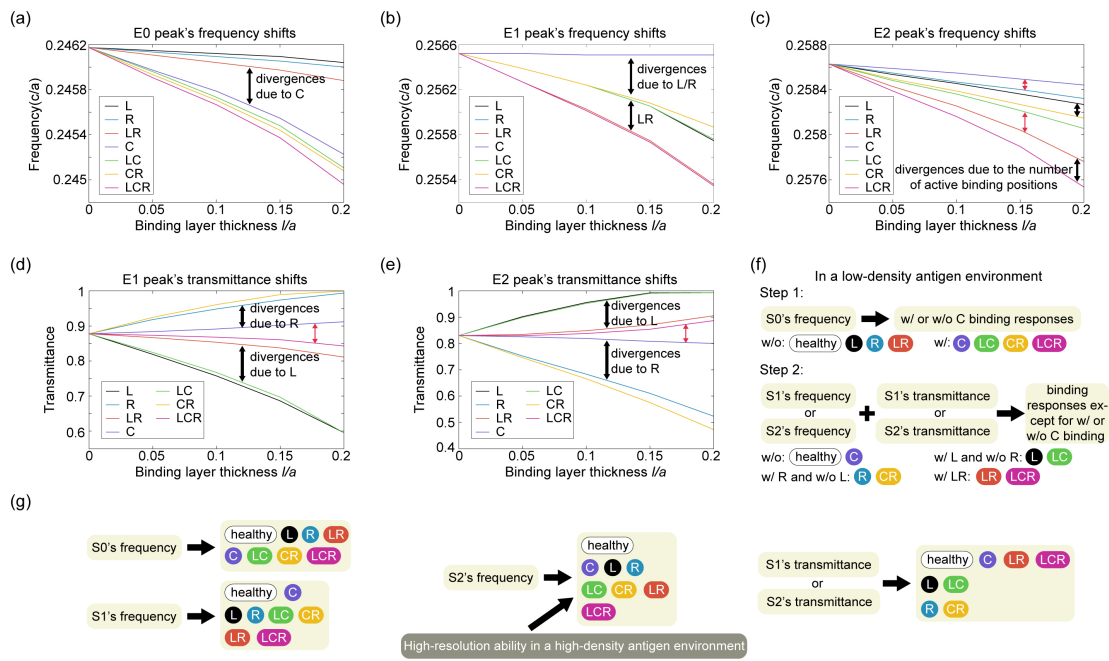

**Figure 5.** Response abilities of five variable observations for seven binding types. (a-c) focus on E0, E1, and E2 peaks' frequency shifts. (d-e) focus on E1 and E2 peaks' transmittance shifts. Seven binding configurations are marked via seven colors shown in the figure legend. Black (red) bidirectional arrows denote the major (minor) divergences. Resolution abilities of all observations are summarized in (g). (f) gives a two-step logic to analyze the binding information in a low-density antigen environment.

**Conclusion and discussion**

Based on the SSH model, we design a lab-on-chip PhC biosensor with topological interface modes. It has been widely validated that they possess strong immunity to disordered disturbances. Thus, differing from the conventional biosensor, our PhC biosensor has a high tolerance for fabrication error. The fundamental architecture of our biosensor is a PhC strip with three interface modes. At least three disease antibodies (or biometric layers) can be placed on these interfaces. Therefore, the simplest single PhC strip can diagnose at least three or seven combined diseases. Moreover, our PhC biosensor can be extended as a quasi-two-dimensional strip array, or, in other words, an integrated diagnostic for multiple diseases.

Another innovation in our work is the asymmetries of our on-chip biosensor. Our PhC architecture is asymmetric, the derived edge modes are asymmetric, and the mode couplings are also asymmetric. Asymmetry implies differentiation, which is beneficial to distinguish the measured signals and improve the resolution ability of our PhC biosensor. Therefore, our biosensing device even has two working modes, which is superior to conventional biosensors. Multi-parameter diagnostic works in a low-density antigen environment, and single-parameter diagnostic works in a high-density antigen environment.

Our biosensor exhibits high sensitivity to multiple disease markers in a single measurement. This multiplexed detection capability is crucial for enhancing the efficiency and reliability of biosensing in practical applications. Our work ushers the development of on-chip PhC biosensors. We believe that the asymmetric design may guide the engineering of future biosensing technologies.

**Appendix: Mode couplings and transmission spectra**

According to the wave functions for E0, E1, and E2 modes shown in the main text Fig. 2(c), we can infer the following Hamiltonian corresponding to the case with $k_y = 0$:

$$H = \begin{pmatrix} \varepsilon_L & t_{LC} & t_{LR} \\ t_{LC} & \varepsilon_C & t_{CR} \\ t_{LR} & t_{LC} & \varepsilon_R \end{pmatrix}. \tag{A1}$$

$\varepsilon_i$ means the eigen frequency of $i$ interface mode. $t_{ij}$ denotes the coupling between $i$ and $j$ interface modes. For our PhC strip setup, the fitting parameters are $\varepsilon_L = 0.2571$, $\varepsilon_C = 0.2485$, $\varepsilon_R = 0.2556$, $t_{LC} = -0.002$, $t_{CR} = 0.004$, and $t_{LR} = -0.0005$ in units of $c/a$. The solved three eigenvalues are $\varepsilon_{E0} = 0.2465c/a$, $\varepsilon_{E1} = 0.2561c/a$, and $\varepsilon_{E2} = 0.2586c/a$, which is close to the frequencies for E0, E1, and E2 modes in main text. And then we picture their eigenvectors in Appx. Fig. 1(a). It is found that they are in good agreement with the major electric fields of three interfaces in Fig. 2(c). Therefore, our fitting parameters are reasonable.

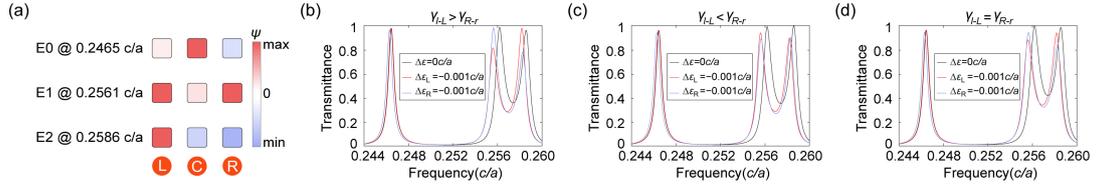

**Figure 6.** (a) Eigen wavefunctions of three-mode coupling mode. (b-d) Calculated transmission spectra via Landauer-Büttiker formulation. In (b), $\gamma_{l-L} = 0.001c/a$, $\gamma_{R-r} = 0.0008c/a$. In (c), $\gamma_{l-L} = 0.0008c/a$, $\gamma_{R-r} = 0.001c/a$. In (d), $\gamma_{l-L} = 0.001c/a$, $\gamma_{R-r} = 0.001c/a$.

To understand the photonic transport through our PhC biosensor, we employ the Landauer-Büttiker theory [1]. In this theory, the transmission spectrum is given by

$$T(\varepsilon) = Tr(\Gamma_L G_r \Gamma_R G_a). \tag{A2}$$

Here, matrix $\Gamma_L$ describes the scattering rates from the photon continuum in left glass to the central PhC strip. Matrix $\Gamma_R$ describes the scattering rates from PhC strip into the photon continuum in the right glass. Green functions $G_r$ and $G_a$ are the retarded and advanced Green functions. And they are connected by the relation, $G_a = G_r^\dagger$. The retarded Geen function takes the following form:

$$G_r = \left[\varepsilon I - H + \frac{i}{2}(\Gamma_L + \Gamma_R)\right]^{-1}. \tag{A3}$$

Since, the central PhC strip can be considered a three-mode coupling model (Eq. A1), scattering matrixes $\Gamma_L$ and $\Gamma_R$ can be expressed by three scattering rates that three modes scatter into the left and right glass. For simplicity, the scattering rates are set as constants: $\gamma_{l-L} = 0.001$, $\gamma_{l-C} = 0.0003$, $\gamma_{l-R} = 0$, $\gamma_{L-r} = 0$, $\gamma_{C-r} = 0.0003$, and $\gamma_{R-r} = 0.0008$ in units of $c/a$. $l$ and $r$ denote the left and right glasses. The special relation, $\gamma_{l-L} > \gamma_{R-r}$, is available. To simulate the accumulation of protein-binding layers in the main text, we reduce the onsite energy of the corresponding modes by $0.001c/a$. And the results are shown in Appx. Fig. 1(b). Since there is a fitting bias, the transmittances of E1 and E2 peaks are higher than those in reference spectrum (main text Fig. 4(a)). However, we can still analyze the effects of binding configurations on the transmittance differences between E1 and E2 peaks. For the case with $\Delta\varepsilon_L = -0.001$, compared to the reference ($\Delta\varepsilon = 0$), one can conclude that L-type binding causes a lower E1 peak and a higher E2 peak, while R-type binding works reversely. These conclusions are consistent with those in the main text. Then, we reverse the scattering rate relation via $\gamma_{l-L} = 0.0008c/a$ and $\gamma_{R-r} = 0.001c/a$. The calculated transmission spectra are shown in Appx. Fig. 1(c). There is no considerable transmittance difference between E1 and E2 peaks. It seems that the influence of asymmetric scattering rates ($\gamma_{l-L} < \gamma_{R-r}$) is restrained. Cases with $\Delta\varepsilon_L = -0.001$ and $\Delta\varepsilon_R = -0.001$ are no longer distinguishable. To find the enforcer of restraint, we set $\gamma_{l-L} = \gamma_{R-r} = 0.001c/a$, and the results are shown in Appx. Fig. 1(d). Obviously, the asymmetry of three-mode coupling model itself causes an effect similar to the occasion dominated by $\gamma_{l-L} > \gamma_{R-r}$, but to a lesser extent. Therefore, the asymmetric scattering rates, mode couplings, and mode hybridizations jointly give rise to the opposite variations on the transmittances of E1 and E2 peaks, and the distinction between L- and R-type protein bindings.

**Appendix reference**